\documentstyle[12pt,aaspp4]{article}

\def\arcdeg {$^{\circ}$}
\def\etal   {{\it et~al.\/}}

\def\sqrdeg  {${\Box^{\circ}}$}

\def\arcsec {$^{\prime \prime}$}
\def\kms    {km\thinspace s$^{-1}$}
\def\HI     {H{\sc i}}

\received{18 February 1999} 
\accepted{17 March 1999} 

\slugcomment{To be published in the {\it Astronomical Journal}, July 1999}

\lefthead{Cabanela and Dickey}
\righthead{Pisces-Perseus Supercluster Spin Vectors}

\begin{document}

\title{Determination of Galaxy Spin Vectors in the Pisces-Perseus Supercluster with the Arecibo Telescope}
\author{J. E. Cabanela and John M. Dickey}
\affil{Astronomy Department, University of Minnesota} 
\affil{116 Church Street SE, Minneapolis, MN 55455}
\authoremail{juan@aps.umn.edu,john@astro.spa.umn.edu}

\begin{abstract} 
We use {\HI} observations made with the upgraded Arecibo 305M Telescope in
August 1998 to obtain accurate spin vector determinations for 54 nearly edge-on
galaxies in the Minnesota Automated Plate Scanner Pisces-Perseus Survey
(MAPS-PP). We introduce a simple observational technique of determining the
sense of rotation for galaxies, even when their {\HI} disks are not fully
resolved.  We examined the spin vector distribution of these 54 galaxies for
evidence of preferential galaxy alignments.  We use the Kuiper statistic, a
variant of the Kolmogorov--Smirnov statistic, to determine the significance of
any anisotropies in the distribution of galaxy spin vectors.  The possibility
of ``spin vector domains'' is also investigated.  We find no significant
evidence of preferential galaxy alignments in this sample.  However, we show that
the small sample size places weak limits on the level of galaxy alignments.
\end{abstract}



\section{Introduction}

The search for galaxy alignments has a long history, beginning with searches
for alignments in ``Spiral and Elliptical Nebulae'' during the late 19th
Century. Recent scrutiny of the problem has been motivated by the understanding
that establishing the level of galaxy spin vector ($\vec{L}$) alignments could
offer an additional constraint on various theories of galaxy formation and
evolution. For example, ``top-down'' scenarios of Large-Scale Structure
formation can lead to ordered distributions of angular momentum on cluster and
supercluster scales through a variety of mechanisms (\cite{zel0}, \cite{dor8},
\cite{whi4}, \cite{col8}). In addition to galaxy $\vec{L}$ alignments resulting
from various formation mechanisms, galaxy $\vec{L}$ alignments may also be the
evolutionary result of anisotropic merger histories (\cite{wes4}),
galaxy-galaxy interactions (\cite{sof2}), or strong gravitational gradients
(\cite{cio4}, \cite{cio8}).  For a summary of the history of the field, see
Djorgovski (1987) and Cabanela \& Aldering (1998) [hereafter Paper I].

Observational support exists for some forms of galaxy $\vec{L}$ alignments with
surrounding large-scale structure.  For example, \cite{bin2} discovered that
the major axes of cD galaxies tended to be aligned with the axes of their parent
cluster.  However, most previous searches for galaxy alignments have had results
that one could describe as negative or statistically significant but not
strongly so.  One complication in earlier efforts has been that most have not
truly determined $\vec{L}$, but rather simply used the position angle (and
sometimes ellipticity) of the galaxies in an attempt to determine the possible
distribution of $\vec{L}$. However, for each combination of galaxy position
angle and ellipticity, there are four solutions for the true orientation of the
galactic angular momentum axis ($\vec{L}$).  This degeneracy in $\vec{L}$ can
only be removed by establishing both which side of the major axis is moving
toward the observer and whether we are viewing the north or south side of the
galaxy, where ``north'' is in the direction the galaxy angular momentum vector.
Therefore previous studies have either restricted themselves to using only
position angles of galaxies, or they have often taken all four possible
solutions of $\vec{L}$ with equal weight (\cite{fli8}, \cite{kas2}).

Several studies have been published regarding searches for alignments using
completely determined galaxy angular momentum axes. \cite{hel2} used {\HI} and
optical observations of 20 galaxies in the Virgo cluster to show that no very
strong $\vec{L}$ alignments exist. However, a followup to this study by
\cite{hel4} found evidence for anti-alignments of spin vectors for binary pairs
of galaxies in a sample of 31 such pairs.  \cite{hof9} briefly investigated the
possibility of galaxy alignments by plotting up the $\vec{L}$ orientations for
$\sim85$ galaxies with fully determined spin vectors from their Virgo cluster
sample and found no obvious alignments.  Most recently, \cite{han5} used a
sample of 60 galaxies from the Third Reference Catalogue of Bright Galaxies
(\cite{deV1}, hereafter referred to as the RC3) in the ``Ursa Major filament''
and found no evidence of galaxy alignments.

There are several criticisms one can level against these earlier studies.  All
the studies attempted to use relatively small samples to map out orientation
preferences over the entire sky.  Thus only very strong $\vec{L}$ alignment
signatures could have been discovered via this method.  The samples were
selected using source catalogs with ``visual'' criteria which may have led to a
biased sample.  For example, as noted in Paper I, the source catalog for the
``Ursa Major filament'' study, the RC3, suffers from the ``diameter-inclination
effect,''  which leads to a strong bias for preferentially including face-on
galaxies over edge-on galaxies of the same diameter (\cite{hui4}).  Finally,
no attempt was made to consider the positions of the galaxies within the local
large-scale structure before looking for alignments.  Considering that the local
mass density is critical for determining which alignment mechanism may be
dominant, an attempt should be made to look for $\vec{L}$ alignments relative to
local large-scale structures.  This study is an attempt to avoid some of the
issues citied above and obtain a sample of galaxies with well determined
$\vec{L}$ in various environments in a supercluster using a
mechanically-selected sample of galaxies.

For this study, we selected a subsample of the Minnesota Automated Plate
Scanner Pisces-Perseus galaxy catalog (hereafter MAPS-PP), which is a true
major-axis diameter-limited catalog built using automated, mechanical methods
and does not exhibit the ``diameter-inclination'' effect (see Paper I).  We
determined the $\vec{L}$ orientation for the galaxies in this subsample using
{\HI} observations.  The sample selection criteria are outlined in Section
\ref{HIobservations}. The analysis methods are discussed in Section
\ref{HIanalysis}. Section \ref{HIresults} discusses the results of the data
analysis. Our interpretation of these results is provided in Section
\ref{HIconclusions}.

\section{Data \label{HIobservations}}
The galaxy sample for this study was selected from the MAPS-PP.  The MAPS-PP
catalog was designed to avoid several of the pitfalls of previous attempts to
measure galaxy orientations.  The MAPS-PP contains $\sim1400$ galaxies in the
Pisces-Perseus Supercluster field with (roughly) isophotal diameter
$>$30{\arcsec} constructed from digitized scans of the blue and red plates of
the Palomar Observatory Sky Survey (POSS I).  By using a mechanical measure of
the diameter, this catalog avoids the ``diameter-inclination'' effect seen in
both the Uppsala General Catalog (\cite{nil4}, hereafter UGC) and the RC3. The
MAPS-PP also uses a two-dimensional, two-component fit of the galaxy light
profile in order to obtain a more accurate position angle and ellipticity
measurement for the component of the galaxy with most of the angular momentum
(e.g. - the disk in spirals). Such a full two-dimensional fit has been shown
(\cite{byu5}) to be very effective at recovering the image parameters
in situations were a simple ellipse fit fails (e.g. - edge-on spirals with a
large bulge).  More details as to the construction of the MAPS-PP are available
in Paper I.

\subsection{Selection Criteria}
For this study, we selected a subsample of the MAPS-PP that could have their
$\vec{L}$ determined through {\HI} observations and at the same time
could probe the galaxy $\vec{L}$ orientations relative to the large-scale
structure of the Pisces-Perseus Supercluster (hereafter PPS). {\HI}
observations can determine which side of the major-axis is approaching us,
reducing the four-fold degeneracy in the $\vec{L}$ to two solutions. However,
because of the great distance to the PPS ($cz \approx 5500~$\kms), the
POSS I images don't generally have enough detail to make out spiral arm
structure, so determining if we were viewing the north or south side of a
galaxy would be difficult without re-imaging the galaxies.  Instead, we choose
to constrain the inclinations of the galaxies in our subsample to be edge-on.
This means we effectively reduce the two-fold degeneracy in $\vec{L}$
solution to a single solution and simultaneously we reduce the galaxy
alignments analysis problem from a full three-dimensional problem to a much
simpler one-dimensional problem.  And because the PPS plane itself is viewed
very close to edge-on (\cite{gio8}), we are simplifying the
problem without losing the ability to probe the angular momentum distribution in
relationship to the PPS plane.  The primary requirement for including a MAPS-PP
galaxy in this study was therefore an ellipticity greater than 0.66.

Other criteria for selecting a MAPS-PP galaxy for our {\HI} program 
were based on observational considerations.  To ensure the galaxy 
could be observed from Arecibo, the Declination was required to be 
less than 36{\arcdeg}.  An O (blue) major-axis diameter between 44{\arcsec} 
and 100{\arcsec} was needed so that the {\HI} disk of the galaxy was 
not too small to be targeted on both sides by the Arecibo beam and not 
too large to be fully sampled.  The galaxy was required to be within 
2.25{\arcdeg} of the PPS midplane (as determined in Paper I) and if 
the redshift was known, it needed to be between $3500-7000~${\kms} in 
order to increase the chances it was a true PPS member.  Finally to 
reduce the sample size, we selected galaxies with O magnitude brighter 
than 17.  This MAPS-PP subsample consisted of 105 galaxies.

The MAPS-PP subsample was cross-identified with the NASA/IPAC Extragalactic
Database (NED) in order to obtain previous radio flux measurements and
redshifts.\footnote{The NASA/IPAC Extragalactic Database (NED) is operated by
the Jet Propulsion Laboratory, California Institute of Technology, under
contract with the National Aeronautics and Space Administration.}  We also
examined the field around each subsample galaxy and eliminated those in crowded
fields, which led to a final MAPS-PP subsample of 96 galaxies (which will
hereafter be referred to as the Arecibo sample), listed in Table \ref{Arecibo_subset}.

\subsection{{\HI} Observations \label{HIobs}}
We obtained 21cm line spectra with the 305m Arecibo telescope of the National
Astronomy and Ionosphere Center over 14 nights between August 6 and August 20, 1998.\footnote{The National Astronomy and Ionosphere Center is operated by 
Cornell University under a cooperative agreement with the National Science 
Foundation.} 
The new Gregorian feed was used with the narrow L band receiver using a
25 MHz bandpass centered on 1394 MHz (1024 channels).  One observation was
performed using a 50 MHz bandpass centered at 1400.5 MHz.  The beamsize of 
the 305m Arecibo dish is approximately 3.3{\arcmin} FWHM.

For each of our Arecibo sample galaxies, we made two sets of ON-OFF
observations, one 90{\arcsec} to the east of the central position along the
major-axis, and a corresponding observation to the west of the galaxy center.
Typically, 5 minute integrations were used for each observation, although some
galaxies were re-observed to allow better measurement of their weak flux and
others known to be bright in {\HI} were observed with shorter integrations.

Preliminary data reduction was performed using ANALYZ at the Arecibo facility.
For each observation, the two polarizations were averaged together.  For each
galaxy we then archived both the sum of the east and west ($E+W$) spectra and the
difference (in the sense east minus west).  It is the difference ($E-W$)
spectra that can be used to determine the spin vector, by allowing us to
determine which side of the major-axis is moving toward us relative to the
galaxy center.  Of the 96 galaxies in the original sample, 6 were 
not observed, 16 were not detected in {\HI}, 3 suffered from strong 
radio frequency interference (RFI), and one suffered from a distorted 
baseline.  We therefore had a total of 70 galaxies for which there 
were good detections.

Subsequent data reduction was performed on the 70 galaxies for which good $E+W$
detections existed.  The spectra were Doppler corrected and the fluxes
corrected for gain differences with zenith angle and changes in system
temperature. A visual estimate of each galaxy's redshift was made and then radio
frequency interference (RFI) within $750~${\kms} of the line was 'removed' from
the spectra.  RFI 'removal' was performed interactively and the RFI was replaced
with a linear interpolation between the two endpoints of the spectra.  Noise was
added to the linear interpolation, using the surrounding spectral channels to
determine the noise level. Both the $E+W$ and $E-W$ spectra were baseline corrected
using a linear fit to non-{\HI} line channels within $500~${\kms}.

We determined the {\HI} line properties of the galaxy using the $E+W$ spectra.
All velocities follow the optical convention, $v=c\Delta\lambda/\lambda_0$, and
are adjusted to be in the heliocentric frame. The flux-weighted mean velocity,
$v_0$, of the galaxy as well as the line flux is computed. The line width used
the mean of the line widths at a threshold of 50\% of the boxcar equivalent
flux and at a threshold of 20\% the maximum flux determined by using a outward
searching algorithm (\cite{lav7}). The reported line width has
been corrected for noise and channel width using the method outlined in Lavezzi
and Dickey (1997).

\subsection{Determination of Galaxy Spin Vector Directions and Uncertainty 
\label{MomentsMethod}}
The direction of the galaxy's spin vector was determined by taking the 
first moment of the $E-W$ spectra, $\mu_{E-W}$, where
\begin{equation} 
\label{EWmom}
\mu_{E-W}~=~\frac{\int_{v_{min}}^{v_{max}} f_{E+W}(v) f_{E-W}(v) (v-v_0) dv}
           {\int_{v_{min}}^{v_{max}} [f_{E+W}(v)]^2  dv}  ,
\end{equation} 
where $v_{min}$ and $v_{max}$ are the minimum and maximum velocity of the line
respectively, $v_0$ is the flux-weighted mean velocity of the galaxy, and
$f_{E+W}(v)$ and $f_{E-W}(v)$ are the fluxes of the $E+W$ and $E-W$ spectra
respectively. Negative $\mu_{E-W}$ implies that the eastern side of the galaxy
is approaching us relative to the galaxy center, meaning the galaxy's $\vec{L}$
points northward.  Positive $\mu_{E-W}$ implies $\vec{L}$ points to the south.

The uncertainty in $\mu_{E-W}$ due to bad baseline and spectral noise was
measured using two variants of the normal first moment.  To determine the
effect of spectral noise on the first moment, we computed $\mu_{offset}$, where
we measure the first moment of the flux outside the line by conserving
$\Delta{v} = (v_{max}-v_{min})$, but offset the $v_0$, $v_{min}$, and $v_{max}$
in equation \ref{EWmom} to lie outside the line (see Figure \ref{mudef}).  This
gave us a measure of the contribution of spectral noise (presumably similar
outside the {\HI} line as inside) to the value of $\mu$.  To determine the
effect of uncertainty in the baseline fit to the first moment determination, we
also computed $\mu_{wide}$, where we find the 1st moment about $v_0$ of the
flux outside the line .  We then scaled this by
${\overline{\Delta{v}}}/{\overline{\Delta{v_{outside}}}}$ to determine the
amount of $\mu_{E-W}$ uncertainty due to uncertainty in the baseline fit.  Both
$\mu_{offset}$ and $\mu_{wide}$ are illustrated in Figure \ref{mudef}. 
$\mu_{offset}$ and $\mu_{wide}$ measurements suggest that galaxies with
$\left|\mu_{E-W}\right|~<~15~${\kms} should be considered to have undetermined
spin (see Figure \ref{fig3-1}). 

To confirm that the $E-W$ spectra are the result of gas being observed on both
sides of the major axis, we also computed the cross-correlation, $P_{cc}$, of
the $E+W$ and $E-W$ spectra, 
\begin{equation} 
\label{EWcc}
P_{cc}~=~\frac{\int_{v_{min}}^{v_{max}} f_{E+W}(v) f_{E-W}(v) dv}
           {\int_{v_{min}}^{v_{max}} [f_{E+W}(v)]^2  dv}  ,
\end{equation} 
since we would expect that the $E+W$ and $E-W$ spectra would be orthogonal in
those cases where the flux is from both the eastern and western positions. We
have empirically found that if $P_{cc}>0.4$ the $E-W$ flux was likely to be
entirely from only one position and thus the spin measurement should be
considered undetermined.  It should be noted that this process will not
eliminate observations of galaxies with an asymmetric {\HI} distribution if
there is significant flux in both the eastern and western positions.   Such a
asymmetric {\HI} distribution would affect the mean velocity, $v_0$, and thus
may affect the amplitude of $\mu_{E-W}$, but it should not change the sign of
$\mu_{E-W}$, which is the observable we use later.

The final dataset had 54 galaxies with well determined spin vectors 
out of the 70 galaxies with good {\HI} detections (see Table
\ref{HI_data}), 16 galaxies having been rejected from the sample due 
to either large $P_{cc}$ or small $\mu_{E-W}$.  For these galaxies, we computed
\begin{equation}
\theta_{\vec{L}} 
=~\theta + 90^{\circ} (\mu_{E-W} / \left| \mu_{E-W} \right|) ,
\end{equation}
which is the projection of $\vec{L}$ on the plane of the sky.  Since the Arecibo
sample is chosen to be nearly edge-on, $\theta_{\vec{L}}$ is essentially a
complete description of $\vec{L}$, allowing simple one-dimensional statistical
analysis to be used for what is normally a three-dimensional problem.

\section{Data Analysis Methods \label{HIanalysis}} 
\subsection{The Kuiper Statistic \label{KuiperStat}}
Identification of anisotropies in the observed $\theta_{\vec{L}}$ and $\theta$
distributions was initially done by using the Kuiper V statistic, which is a
two-sided variant of the Kolmogorov-Smirnov (K--S) D statistic (\cite{pre2}).
We use the Kuiper V statistic because the K--S D statistic can systematically
underestimate the significance of differences between the observations and the
models, especially if the differences are near the ends of the distribution
(\cite{pre2}). For this test, we compare the cumulative distributions of a variable,
$x$ (such as $\theta_{\vec{L}}$, $\Delta\theta_{\vec{L}}$, etc.), in the
observed sample, $S(x)$, with that for a model of 100000 randomly-oriented
galaxies, $S_m(x)$. The Kuiper statistic, $V$, is then defined as
\begin{equation} 
\label{KuiperDef}
V~=~D_{+} + D_{-}~=~ max [S(x)-S_m(x)] + max [S_m(x)-S(x)] ,
\end{equation} 
the sum of the absolute values of the maximum positive ($D_{+}$) and negative
($D_{-}$) differences between $S(x)$ and $S_m(x)$.\footnote{Note that the
normal K--S D statistic is equal to $max |S(x)-S_m(x)|$.  It doesn't
distinguish between differences above or below the $S_m(x)$ the curve.}   $V$
is essentially a measure of the difference between two distributions (see
Figure \ref{VDef}).  If the number of degrees of freedom is known a priori, a simple
functional form exists for the probability, $P(V)$, that the two samples whose
cumulative distributions differ by $V$ were drawn from the same parent
distribution (see Press {\etal} 1992, for example). Therefore, if we are
comparing the distribution of $x$ for the observed sample to that of a modeled,
randomly-oriented sample, we have a way of estimating the probability that the
observed sample is drawn from an isotropic distribution. In this study, we
considered a distribution's anisotropy significant if the probability, $P(V)$,
that the Arecibo sample could have been drawn from the randomly-oriented sample
was less than 5\%.

In those cases where the number of degrees of freedom is not well determined a
priori, we used Monte Carlo comparisons of the observations with 1000 model
samples of equal size.  This was necessary in order to avoid overestimating the
significance of an observed anisotropy.  We model a randomly oriented
distribution of galaxies by taking the observed sample, randomly reassigning
the observed $P_{cc}$ and $\mu$ values to various galaxies ($\mu$ is determined
by randomly reversing the sign of $|\mu|$), and then randomly generating the
major-axis position angle, $\theta$.  This model kept the spatial distribution
of the original sample and the {\HI} observational selection effects while
otherwise being a completely randomly oriented model. Comparison of the real
distribution of a variable versus its distribution in the 1000 Monte-Carlo
samples is used to determine the significance of an anisotropy in some of the
more complicated distributions discussed in section \ref{HIresults}.

\section{Results and Analysis \label{HIresults}}

\subsection{Probing for Global Spin Vector Alignments \label{followup}}

As a followup to the work done in Paper I, we initially examined some of the
distributions similar in nature to the ones investigated that study.  We
divided the entire MAPS-PP and Arecibo Samples into 3 subsets each: the high
density subset, the low density subset, and the complete sample.  The high and
low density subsets were created using surface density estimates, $\Sigma$,
from the MAPS-PP catalog to compute the median surface density.  The high and
low density subsets include all galaxies with $\Sigma$ greater than and less
than this median value, respectively.  For the MAPS-PP subsets we tested the
$\theta$-based distributions, whereas for the Arecibo subsets, we tested the
$\theta_{\vec{L}}$-based distributions. Examinations of the $\theta_{\vec{L}}$
and $\theta$ distributions show no significant anisotropy in any of the Arecibo
or MAPS-PP subsets.  Similar results were seen for distributions of
$\theta_{\vec{L}}$ and $\theta$ relative to other critical angles including the
following:
\begin{itemize}
\item $\Delta\theta_{\vec{L}}(1)$ and $\Delta\theta(1)$: the difference  of
$\theta_{\vec{L}}$ and $\theta$, respectively, between nearest neighbor
galaxies in that sample.  Note that $\Delta\theta(1)$ is used in the Arecibo
sample only to separate the significance of any $\Delta\theta_{\vec{L}}(1)$
alignments from any $\Delta\theta(1)$ alignments. 
\item $\Delta\theta_{\vec{L}}(Geo)$: the difference of $\theta_{\vec{L}}$ from
the geodesic to the nearest neighbor galaxy.
\item $\Delta\theta_{\vec{L}}(Ridge)$: the difference of $\theta_{\vec{L}}$
from angle of the Pisces-Perseus Supercluster ridgeline at its nearest point
(as determined in Paper I).
\item $\Delta\theta_{\vec{L}}(GC-X)$: the difference of $\theta_{\vec{L}}$ from 
the galaxy concentration position angle built using a percolation length of $X$ 
arcminutes (galaxy concentrations groupings of galaxies identified using a 2 
dimensional friends-of-friends algorithm (redshift is ignored), see Paper I 
for details).
\item $\Delta\theta_{\vec{L}}(GCR-X)$: the difference of $\theta_{\vec{L}}$ from
the radial line to the center of the galaxy concentration built using a
percolation length of $X$ arcminutes.
\end{itemize}

These results, shown in Table \ref{LvecTests}, support the observations in Paper
I in that no simple $\theta$ or $\theta_{\vec{L}}$ alignments appear to be
present. Examination of the $\Delta\theta_{\vec{L}}(GC-X)$ distribution does not
support the tentative anti-alignments seen in Paper I.  We looked for 'twisting'
of $\Delta\theta_{\vec{L}}(Ridge)$ versus distance from the PPS ridgeline, and
could not corroborate this signal seen in the $\Delta\theta(Ridge)$ distribution
of the MAPS-PP in Paper I.  We note that the Arecibo sample is considerably
smaller that the MAPS-PP, so we cannot rule out the trends seen in Paper I, but
we simply cannot support them.

\subsection{Probing for Spin Vector Domains \label{domains}}

An initial visual inspection of the plot of the distribution of
$\theta_{\vec{L}}$ on the sky (Figure \ref{lvecmap}) appears to show some
$\theta_{\vec{L}}$ alignments.  Specifically, in many cases if you pick a galaxy
at random and then compare its $\theta_{\vec{L}}$ with that of nearby galaxies,
the difference is often less than 90{\arcdeg}.  It appeared to the authors that
there was a visual impression of the PPS being divided up into ``spin vector
domains,'' regions with preferred $\vec{L}$ orientations. Because visual
impressions are subjective, we devised tests to look for possible spin vector
domains as well as looking for the alignments of the sort reported in Paper I
for the galaxy major-axes.

We attempted to confirm visual impression of $\vec{L}$ domains seen in Figure
\ref{lvecmap} by examining the orientations of several nearest neighbors,
instead of just the nearest neighbor.  To this end, we computed the
$\Delta\theta_{\vec{L}}(N)$ distribution, which is the summed distribution of
$\Delta\theta_{\vec{L}}$ (respectively) for the N closest galaxies within
3{\arcdeg} of each galaxy . If $\vec{L}$ domains exist, the
$\Delta\theta_{\vec{L}}(N)$ distribution should be peaked toward the lower values
of $\Delta\theta_{\vec{L}}(N)$.

Because the $\Delta\theta_{\vec{L}}(N)$ distribution about one galaxy is not
independent of the distribution about that galaxy's nearest neighbors, the
number of degrees of freedom is uncertain {\it a priori}. This means that the
standard function to determine the probability, $P(V)$, of two distributions
being identical doesn't work.  Instead, we gauge $P(V)$ by generating 1000
Monte Carlo samples and computing the Kuiper V statistic of their
$\Delta\theta_{\vec{L}}(N)$ distributions.  By comparing the value of V for the
observed sample with the distribution of V in the 1000 Monte Carlo samples, we
have an estimate of the likelihood that a greater value of V  is obtained, $P(V)$. 
We therefore use $P(>V)$ in leu of the $P(V)$ used in cases where we know the 
number of degrees of freedom.

We examined the $\Delta\theta_{\vec{L}}(N)$ distributions for the N closest
galaxies of the Arecibo samples, for N ranging from 3 to 10.  These samples
show no significant anisotropy when compared to Monte Carlo generated datasets,
indicating that the visual impression of $\vec{L}$ domains is either incorrect,
or the $\vec{L}$ domains are too weakly aligned to confirm with this test.

Because in Paper I only a simple nearest neighbor test was performed, we also
examined the $\Delta\theta(N)$ distribution for the MAPS-PP samples, in order to
see if $\vec{L}$ domains might be visible in the larger MAPS-PP dataset. We
found that for N ranging from 3 to 10, the $\Delta\theta(N)$ distributions
showed no evidence of significant anisotropies.  This appears to indicate that it
is unlikely that $\vec{L}$ domains exist in the Pisces-Perseus Supercluster.

\subsection{Establishing Limits on Galaxy Alignments \label{limits}}
In order to quantify the largest anisotropic signature that could remain
``hidden'' from our statistical techniques, we performed a simple simulation. We
generated samples drawn from random `sinusoidal' distributions described by
the probability distributions 
\begin{equation} 
\label{cos_prob}
P(\Theta) d\Theta~=~\left[1+\alpha\cos\left(\Theta \frac{2\pi}{\lambda}\right)\right] d\Theta, 
\mbox{where}~\Theta \in \left[0,\lambda\right],
\end{equation} 
and
\begin{equation} 
\label{cos_prob_2}
P(\Theta) d\Theta~=~\left[1+\alpha\cos\left(\Theta \frac{2\pi}{\lambda}\right)\right] d\Theta, 
\mbox{where}~\Theta \in \left[0,\frac{\lambda}{2}\right],
\end{equation} 
where $\alpha$ is the amplitude of the 'sinusoidal' component of the probability
in percent.  In these two distributions, $\Theta$ represents either the expected
$\theta$ or $\theta_{\vec{L}}$ distributions in the cases of large-scale
alignments (equation \ref{cos_prob_2}), or the $\Delta\theta$ and
$\Delta\theta_{\vec{L}}$ distributions in the cases of alignments (equation
\ref{cos_prob_2}), anti-alignments (equation \ref{cos_prob_2}) or both (equation
\ref{cos_prob}) between nearby galaxies.  Using these two distributions, we can
generate samples with a predetermined amplitude, $\alpha$, for the alignments
present and then compute the value of $P(V)$, the probability of the sample
having been drawn from a random sample. By repeatedly doing this, we can
determine the distribution of $P(V)$ for a given $\alpha$ and sample size.

For samples of 30, 54, 100, 615, and 1230 galaxies (the sizes of our subsets as
noted in Table \ref{LvecTests}), we computed the $P(V)$ distribution for 100
generated $\Theta$ samples with amplitudes, $\alpha$, ranging from 0\% to 100\%
in steps of two percent (see Figure \ref{ampVprobs}). We then examined at which
point 95\% of the $P(V)$ distribution dropped below 0.05, our criterion for
calling a distribution significantly anisotropic. This gave us an estimate of the
largest amplitude sinusoidal anisotropy that could have been missed, which we
call $\alpha_{95}$.  $\alpha_{95}$ is therefore the smallest amplitude of a
sinusoidal anisotropy for which there is a 95\% chance of detection given the
criteria of $P(V) < 0.05$.

For our Arecibo sample, we find that with 54 galaxies $\alpha_{95} \approx
0.75$, therefore we can only eliminate global spin vector alignments with
sinusoidal amplitudes greater than 75\%.  This sample does not place very
strong limits on level of any spin vector alignments present. With the 1230
galaxies in the MAPS-PP catalog, we find $\alpha_{95} \approx 0.15$.  Therefore
we can eliminate the possibility of galaxy major-axis alignments at amplitudes
greater than 15\%.  Major-axis alignments place very weak limits on the level
of spin vector alignments due to the fact that the orientation of the
major-axis of the galaxy, with no additional information, only restricts the
spin vector to a plane.  However, if there is a spin vector alignment, it must
be reflected in the major-axis distribution of the edge-on galaxies.  We find
that in a subsample of 729 MAPS-PP galaxies restricted to $\epsilon>0.50$, 
there is no significant major-axis anisotropy of any sort.  For a sample of 729
galaxies, we find $\alpha_{95} \approx 0.20$, therefore, we can confidently state
that there are no spin vector alignments with sinusoidal amplitude greater than 
20\% (within the uncertainty due to the two-fold degeneracy in mapping
major-axis position angle to spin vector).

We would like to have computed $\alpha_{95}$ for the spin vector domain tests
in order to gauge their sensitivity but it was computationally too expensive.

\section{Conclusions \label{HIconclusions}}

We have constructed the only catalog of well determined spin vectors for
galaxies in the Pisces-Perseus Supercluster.  Our study is the first radio study
that explicitly looks at the spin vector distribution of galaxies
in a supercluster and was optimized toward that end.  We developed a simple
technique for obtaining spin vector determinations and accessing the level of
uncertainty in the spin vector determinations due to both spectral noise and
uncertainty in fitting the continuum.  We were intentionally rather conservative
in our data selection criteria, possibly rejecting several well measured spin
vectors.  

There are several problems currently hampering the determination of the angular
momentum distribution of galaxies relative to each other and to the surrounding
large-scale structure.  One major problem is that we do not have a very clear
understanding of the internal extinction in galaxies and its effect on the
appearance of the galaxy with changing inclination.  Therefore, it is very
difficult to accurately determine the inclination of a galaxy based solely on
its ellipticity and position angle.  This also makes it more difficult to
construct a proper volume-limited sample for a large-scale angular momentum
study.  One could obtain redshifts for all the galaxies in a diameter-limited
or magnitude-limited galaxy catalog and select a volume-limited subsample, but
without a clear understanding of internal extinction, we cannot correct
magnitudes and diameters for inclination.

We compensated for these uncertainties of the effects of internal galaxy
extinction by restricting our sample to highly edge-on galaxies.  This had the
added benefit of making the {\HI} spectra of the galaxies as broad as possible,
and thus making it easier to determine the $\vec{L}$ orientation.  We note that
this restriction to edge-ons could make reduction of alignments relative to
large-scale structure difficult, since we would be restricting analysis to
galaxies with $\vec{L}$ in the plane of the sky.  However, in this study, the
edge-on orientation of the Pisces-Perseus Supercluster means our sample 
galaxies' $\vec{L}$ lie in the plane perpendicular to the supercluster plane, 
which is advantageous for reducing the complexity of the analysis.  This does
reduce our sensitivity to any $\vec{L}$ alignments that lie outside the plane 
of the sky.  For example, if galaxies' $\vec{L}$ are preferentially oriented 
in a given direction within the plane of the Pisces-Perseus Supercluster (e.g. 
toward a cluster in the supercluster plane) rather than simply being restricted
to that plane, we may not detect such an alignment in our sample, since we 
restrict $\vec{L}$ of sample galaxies to the plane perpendicular to the 
supercluster plane.  It would be interesting to perform similar observations 
of a ``face-on'' version of Pisces-Perseus, where we would then be restricting 
$\vec{L}$ to the supercluster plane and possibly investigating a new class of 
$\vec{L}$ alignments.

The technique we outline for obtaining spin vector measurements could be
applied to quickly obtain $\vec{L}$ measurements for many galaxies in
superclusters other than Pisces-Perseus.  It is also notable that this
technique could be transferred to multi-fiber spectroscopy.  By assigning two
fibers to each galaxy, one could simultaneously determine the $\vec{L}$
directions of many galaxies much more quickly than a comparable line
slit spectrograph observations.  No rotation curve information would be
available, but it would allow quick collection of a large sample of well
determined galaxy $\vec{L}$. 

Our examination of the $\vec{L}$ distribution of galaxies in Pisces-Perseus
provides no support for any form of anisotropic $\vec{L}$ distribution. We do
not provide confirmation of the possible $\vec{L}$ alignments noted in Paper I
for the major-axis distributions of Pisces-Perseus galaxies. Given the
relatively small size of the Arecibo sample, rather large anisotropies in the
spin vector distribution of the Arecibo sample (see Section \ref{limits}) could
remain undetected with our technique.  We do note that by using a sample of 729
nearly edge-on galaxies from the original MAPS-PP catalog, we feel we can
restrict the sinusoidal amplitude of any spin vector anisotropy present to be
less than approximately 20\% the background `random' distribution, at least in
the plane perpendicular to the Pisces-Perseus supercluster ridge.

It is unclear at what level galaxy $\vec{L}$ alignments might be expected as no
recent simulations have been designed with the goal of estimating galaxy
alignments. We expect that if galaxy alignments are produced by large-scale
structure formation, the alignments would be strongest in areas of low density,
where the relative scarcity of subsequent galaxy-galaxy interactions suggests
the initial $\vec{L}$ distribution would be better preserved. However, 
as noted in the introduction, galaxy alignments can arise from a variety of 
evolutionary processes, in both high and low density environments. It would be 
interesting if in modern computer simulations of galaxy evolution, the angular 
momentum of the resulting galaxies was investigated for $\vec{L}$ alignments 
and predictions as to the amplitude (and type) of any anisotropies in the 
$\vec{L}$ distribution were made.

As we showed in Section \ref{limits}, sample sizes need to be large (on the
order of at least 500 galaxies) in order to unambiguously detect weak
alignments. There are two paths toward increasing the sample size.  We could
examine a denser cluster with a greater number of targets satisfying our
edge-on criteria such as the Coma cluster.  It would be interesting to
investigate the possibility of tidally induced galaxy alignments in denser
environments as predicted by Ciotti and Dutta (1994) and Ciotti and Giampieri
(1998).  The only previous study looking for galaxy alignments in Coma was
plagued by stretched imaging (\cite{djo7}), so alignment results for this
cluster are still unclear.  Our other option for increasing sample size is to
develop a better understanding of the internal extinction in galaxies so that
we could use galaxies of all inclinations.  The first author is currently
investigating using image parameters of a large number of galaxies obtained
using the APS database in order to better determine the internal extinction
properties of galaxies.

We would like to thank telescope operators Miguel Boggiano, Willie Portalatin,
Pedro Torres, and Norberto Despiau for their good humor and help with observing
(And especially Norberto for his ``lucky coffee''). JEC would like to thank
Chris Salter, Tapasi Ghosh, Jo Ann Etter, and Phil Perillat for helping make his
first radio observing experience excellent, both professionally and personally.
Travel was sponsored by the National Astronomy and Ionosphere Center (NAIC) and
the University of Minnesota Graduate School.

This research has made use of the APS Catalog of the POSS I, which is supported
by the National Aeronautics and Space Administration and the University of
Minnesota. The APS databases can be accessed at {\it http://aps.umn.edu/} on the
World Wide Web.  Some data reduction was performed at the Laboratory for
Computational Science and Engineering (LCSE) at the University of Minnesota. 
Information about the LCSE can be found online at {\it http://www.lcse.umn.edu/}.

\clearpage
\begin{deluxetable}{llccccccc} 
\tablecaption{The Arecibo Sample (A subset of 96 MAPS-PP galaxies) \label{Arecibo_subset}}
\scriptsize
\tablenum{1}
\tablewidth{0pt} 
\tablehead{APS ID & Common Name\tablenotemark{a} & $\alpha$ & $\delta$ & $m_{O}$\tablenotemark{b} & $O-E$\tablenotemark{b} & $a_{O}$\tablenotemark{b} & $R_{ridge}$\tablenotemark{c} & $\Sigma$\tablenotemark{c} \\
 & & (B1950) & (B1950) & & & & (\arcdeg) & }
\tablenotetext{a}{The common name of the object was determined via cross-identification of APS position with the NASA/IPAC Extragalactic Database (NED).}
\tablenotetext{b}{The O bandpass magnitude ($m_O$), $O-E$ color, and O bandpass major-axis diameter ($a_O$) were all obtained from the APS catalog.  The $m_O$ and $O-E$ are zeropointed on a plate-by-plate basis as outlined in Cabanela \& Aldering (1998).}
\tablenotetext{c}{The distance from the Pisces-Perseus ridgeline ($R_{ridge}$) and local surface density ($\Sigma$) are from Cabanela \& Aldering (1998). $\Sigma$ is in units of galaxies/\sqrdeg.}
\startdata
O\_778\_873376 & UGC 11993 & 22:18:25.5 & 34:58:14.8 & 15.67 & 1.49 & 61.7 & 0.89 & 5.60 \nl 
O\_778\_731211 & 22208+3548 & 22:20:49.5 & 35:48:04.8 & 14.66 & 1.42 & 52.0 & 0.25 & 25.35 \nl 
O\_778\_700353 & 22233+3556 & 22:23:19.2 & 35:55:31.0 & 15.32 & 1.52 & 57.4 & 0.58 & 10.40 \nl 
O\_778\_847676 & None & 22:24:45.4 & 35:10:40.3 & 15.41 & 1.39 & 44.0 & 0.27 & 11.23 \nl 
O\_778\_849054 & None & 22:25:53.3 & 35:04:31.8 & 15.39 & 1.66 & 46.7 & 0.12 & 9.07 \nl 
O\_778\_1040599 & NGC 7320B & 22:35:10.3 & 33:39:46.9 & 15.54 & 1.47 & 49.2 & 0.34 & 7.17 \nl 
O\_778\_923367 & None & 22:39:36.3 & 34:39:30.7 & 14.03 & 1.39 & 94.8 & 1.12 & 6.39 \nl 
O\_778\_755586 & None & 22:41:32.8 & 35:38:55.2 & 15.39 & 1.52 & 52.1 & 2.19 & 3.20 \nl 
O\_1184\_28270 & None & 22:50:24.2 & 33:07:53.6 & 16.26 & 1.64 & 44.3 & 0.86 & 3.11 \nl 
O\_1184\_66567 & 22508+3230 & 22:50:47.0 & 32:29:50.0 & 14.99 & 1.44 & 49.5 & 0.32 & 11.28 \nl 
O\_1184\_128370 & UGC 12231 & 22:51:11.0 & 31:21:11.7 & 14.84 & -0.35 & 62.4 & 0.84 & 10.07 \nl 
O\_1184\_81567 & None & 22:55:40.5 & 32:11:55.6 & 15.90 & 1.34 & 49.2 & 0.12 & 6.26 \nl 
O\_1184\_196807 & UGC 12320 & 22:59:40.4 & 30:29:42.0 & 15.83 & 1.45 & 66.8 & 1.50 & 6.00 \nl 
O\_1184\_121956 & UGC 12362 & 23:03:51.4 & 31:36:49.0 & 14.44 & 1.00 & 67.4 & 0.17 & 2.27 \nl 
O\_1184\_275037 & MCG +05-54-039 & 23:07:04.4 & 29:12:43.0 & 15.38 & 1.16 & 53.0 & 1.43 & 2.35 \nl 
O\_1184\_313727 & UGC 12427 & 23:10:57.6 & 28:40:55.9 & 14.94 & 0.19 & 49.4 & 0.97 & 5.94 \nl 
O\_1184\_347214 & None & 23:12:38.9 & 28:00:36.0 & 15.60 & 0.46 & 52.7 & 0.99 & 6.24 \nl 
O\_1184\_189398 & UGC 12458 & 23:12:43.0 & 30:40:27.0 & 16.22 & 1.51 & 59.9 & 0.52 & 6.82 \nl 
O\_843\_144830 & None & 23:18:38.9 & 25:22:01.9 & 15.86 & 1.50 & 44.0 & 2.07 & 4.80 \nl 
O\_843\_65466 & None & 23:18:59.9 & 26:12:15.6 & 15.95 & 0.96 & 44.9 & 1.31 & 4.95 \nl 
O\_914\_404205 & UGC 12557 & 23:20:01.1 & 28:54:24.1 & 14.72 & 1.49 & 75.0 & 0.86 & 5.05 \nl 
O\_914\_406137 & None & 23:23:49.0 & 28:58:48.0 & 16.70 & 1.46 & 44.8 & 1.60 & 5.66 \nl 
O\_914\_344933 & UGC 12625 & 23:26:38.5 & 29:29:58.3 & 14.73 & 1.57 & 84.2 & 1.72 & 4.36 \nl 
O\_914\_409641 & UGC 12644 & 23:28:58.8 & 28:54:47.7 & 14.86 & 0.56 & 59.1 & 0.95 & 3.69 \nl 
O\_914\_371708 & None & 23:32:07.5 & 29:26:30.5 & 16.45 & 1.45 & 48.7 & 0.89 & 3.28 \nl 
O\_914\_413061 & None & 23:32:17.4 & 29:02:11.2 & 16.07 & 1.47 & 45.5 & 0.59 & 3.05 \nl 
O\_914\_511814 & None & 23:33:55.7 & 27:39:32.0 & 16.07 & 1.24 & 55.7 & 0.72 & 3.32 \nl 
O\_914\_416774 & UGC 12730 & 23:38:03.3 & 28:54:39.3 & 14.29 & 1.56 & 98.6 & 0.32 & 2.84 \nl 
O\_914\_286281 & UGC 12741 & 23:39:25.1 & 30:18:15.5 & 14.54 & 0.94 & 57.3 & 0.74 & 1.49 \nl 
O\_914\_286479 & None & 23:39:59.4 & 30:19:12.6 & 15.63 & 0.68 & 44.0 & 0.72 & 1.50 \nl 
O\_914\_514191 & None & 23:40:00.1 & 27:46:03.4 & 16.26 & 1.17 & 52.8 & 1.53 & 6.24 \nl 
O\_914\_437214 & CGCG 498-006 & 23:42:35.0 & 28:47:11.1 & 15.55 & 1.38 & 46.1 & 0.93 & 4.75 \nl 
O\_1257\_181710 & MCG +05-01-003 & 23:52:48.7 & 30:06:24.8 & 15.11 & 1.51 & 54.2 & 0.27 & 3.38 \nl 
O\_1257\_106449 & UGC 12845 & 23:53:09.1 & 31:37:15.8 & 15.11 & 1.99 & 52.4 & 1.25 & 2.20 \nl 
O\_1257\_149828 & UGC 12864 & 23:54:50.8 & 30:42:49.4 & 14.49 & 1.11 & 85.0 & 0.34 & 3.85 \nl 
O\_1257\_140283 & NGC 7799 & 23:56:46.8 & 31:00:22.1 & 15.96 & 1.19 & 66.9 & 0.64 & 2.90 \nl 
O\_1257\_307025 & UGC 124 & 00:10:48.1 & 28:05:27.8 & 15.45 & 1.85 & 55.1 & 2.12 & 5.95 \nl 
O\_1257\_224112 & UGC 147 & 00:13:07.5 & 29:23:21.2 & 15.39 & 1.83 & 45.6 & 0.79 & 7.56 \nl 
O\_1257\_212633 & 00139+2939 & 00:13:52.2 & 29:38:45.5 & 15.37 & 0.78 & 67.6 & 0.53 & 18.31 \nl 
O\_1244\_265500 & UGC 238 & 00:22:25.5 & 31:03:58.5 & 13.70 & 1.15 & 99.0 & 0.99 & 3.82 \nl 
O\_1244\_376417 & 00254+3029 & 00:25:21.3 & 30:29:18.9 & 15.20 & 1.34 & 55.9 & 0.48 & 5.08 \nl 
O\_1244\_340721 & UGC 279 & 00:25:36.7 & 30:31:37.5 & 13.90 & 1.09 & 92.3 & 0.51 & 4.61 \nl 
O\_1244\_270335 & 00267+3106 & 00:26:45.2 & 31:06:17.2 & 15.57 & 1.07 & 83.8 & 1.12 & 6.73 \nl 
O\_1244\_679996 & UGC 310 & 00:28:39.0 & 28:42:58.5 & 14.84 & 0.62 & 68.2 & 1.23 & 2.29 \nl 
O\_1244\_578706 & None & 00:29:14.4 & 29:25:41.8 & 15.88 & 0.74 & 45.7 & 0.54 & 3.48 \nl 
O\_1244\_241809 & 00313+3110 & 00:31:15.4 & 31:10:30.6 & 15.40 & 0.00 & 55.4 & 1.30 & 4.67 \nl 
O\_1244\_767827 & UGC 345 & 00:32:09.7 & 28:08:02.5 & 15.22 & 0.09 & 76.1 & 1.75 & 1.76 \nl 
O\_1244\_185805 & 00333+3136 & 00:33:15.9 & 31:35:41.4 & 14.23 & 1.24 & 68.0 & 1.75 & 3.54 \nl 
O\_1244\_655424 & 00347+2853 & 00:34:41.6 & 28:52:27.1 & 14.85 & 0.84 & 98.4 & 0.95 & 7.96 \nl 
O\_1244\_554203 & UGC 412 & 00:36:49.5 & 29:29:16.8 & 15.25 & 1.29 & 55.9 & 0.34 & 10.90 \nl 
\tablebreak
O\_1244\_595079 & UGC 449 & 00:39:41.7 & 29:25:26.2 & 14.96 & 0.93 & 63.5 & 0.40 & 7.98 \nl 
O\_1244\_521380 & None & 00:40:48.3 & 29:46:45.6 & 15.32 & 0.75 & 51.2 & 0.04 & 12.24 \nl 
O\_1244\_442897 & None & 00:41:53.2 & 30:04:51.5 & 16.59 & 0.50 & 47.1 & 0.42 & 11.42 \nl 
O\_1244\_484644 & UGC 478 & 00:43:44.5 & 29:57:54.9 & 14.45 & 1.13 & 80.4 & 0.17 & 13.26 \nl 
O\_601\_2598615 & UGC 501 & 00:46:21.1 & 27:56:44.1 & 15.47 & 1.55 & 91.9 & 1.95 & 2.21 \nl 
O\_601\_927741 & UGC 511 & 00:47:27.3 & 31:27:32.8 & 15.46 & 0.96 & 69.8 & 1.47 & 5.16 \nl 
O\_601\_2448395 & CGCG 501-024 & 00:48:09.5 & 28:25:40.3 & 15.69 & 1.19 & 44.6 & 1.50 & 4.85 \nl 
O\_601\_1985337 & UGC 525 & 00:48:52.4 & 29:26:42.9 & 15.69 & 1.42 & 73.9 & 0.64 & 4.96 \nl 
O\_601\_1986315 & 00494+2924 & 00:49:30.1 & 29:24:19.4 & 14.42 & 1.67 & 60.4 & 0.66 & 6.53 \nl 
O\_601\_2601958 & None & 00:50:49.0 & 28:00:21.8 & 16.80 & 1.60 & 53.5 & 2.08 & 3.56 \nl 
O\_601\_1152906 & UGC 557 & 00:52:03.8 & 31:05:40.0 & 14.56 & 0.73 & 59.5 & 0.76 & 20.54 \nl 
O\_601\_2454111 & UGC 554 & 00:52:04.6 & 28:26:44.9 & 15.22 & 1.79 & 50.9 & 1.71 & 5.65 \nl 
O\_601\_2363374 & 00521+2835 & 00:52:06.5 & 28:35:46.1 & 14.78 & 1.79 & 58.7 & 1.57 & 4.48 \nl 
O\_601\_1044227 & UGC 565 & 00:52:38.6 & 31:24:14.1 & 15.13 & 0.98 & 48.1 & 0.92 & 13.23 \nl 
O\_601\_1267760 & CGCG 501-048 & 00:53:26.0 & 30:48:15.4 & 15.38 & 1.21 & 55.5 & 0.47 & 9.02 \nl 
O\_601\_1050939 & UGC 598 & 00:55:06.2 & 31:12:52.1 & 14.19 & 1.44 & 71.3 & 0.49 & 6.20 \nl 
O\_601\_1498038 & UGC 624 & 00:57:52.7 & 30:23:58.1 & 13.00 & 0.96 & 99.9 & 0.60 & 6.63 \nl 
O\_601\_1063679 & UGC 633 & 00:58:37.0 & 31:14:23.6 & 14.16 & 0.95 & 90.7 & 0.01 & 6.20 \nl 
O\_601\_863271 & None & 01:00:55.7 & 31:46:46.0 & 15.29 & 1.04 & 60.7 & 0.15 & 8.98 \nl 
O\_601\_1294945 & 01011+3056 & 01:01:04.3 & 30:55:47.2 & 15.71 & 1.11 & 57.4 & 0.61 & 4.65 \nl 
O\_601\_978154 & UGC 669 & 01:02:34.0 & 31:24:52.0 & 15.19 & 1.32 & 72.1 & 0.39 & 6.37 \nl 
O\_601\_1197694 & UGC 673 & 01:03:24.6 & 31:08:18.2 & 15.15 & 0.81 & 57.1 & 0.72 & 4.65 \nl 
O\_601\_657010 & UGC 679 & 01:04:17.9 & 32:07:21.1 & 16.17 & 0.63 & 51.3 & 0.10 & 22.35 \nl 
O\_601\_992015 & CGCG 501-092 & 01:05:17.4 & 31:24:29.8 & 15.25 & 1.34 & 45.1 & 0.80 & 9.63 \nl 
O\_601\_175820 & A82-91 & 01:05:18.4 & 33:11:11.4 & 14.49 & 1.25 & 57.8 & 0.82 & 17.03 \nl 
O\_601\_264028 & NGC 407 & 01:07:49.8 & 32:51:38.8 & 14.09 & 1.44 & 96.5 & 0.34 & 18.31 \nl 
O\_601\_359427 & None & 01:07:57.9 & 32:45:57.2 & 16.27 & 0.63 & 47.3 & 0.25 & 19.90 \nl 
O\_406\_436424 & 01104+3443 & 01:10:24.4 & 34:42:01.4 & 16.52 & 0.66 & 49.0 & 1.94 & 2.14 \nl 
O\_406\_502375 & UGC 809 & 01:13:04.0 & 33:32:50.3 & 14.84 & 0.74 & 74.2 & 0.68 & 2.66 \nl 
O\_1189\_285025 & None & 01:19:39.9 & 33:46:42.9 & 15.99 & 0.83 & 45.8 & 0.69 & 8.12 \nl 
O\_1189\_293769 & NGC 512 & 01:21:10.7 & 33:38:47.5 & 13.74 & 1.05 & 69.5 & 0.49 & 6.95 \nl 
O\_1189\_251296 & 01287+3432 & 01:28:43.2 & 34:31:32.6 & 14.81 & 0.55 & 56.1 & 0.76 & 3.02 \nl 
O\_1189\_224928 & NGC 634 & 01:35:25.4 & 35:06:38.8 & 13.70 & 1.13 & 84.6 & 0.57 & 3.98 \nl 
O\_1189\_244311 & UGC 1166 & 01:35:42.0 & 34:44:18.4 & 14.17 & 1.23 & 71.1 & 0.39 & 3.38 \nl 
O\_1189\_234487 & 01366+3455 & 01:36:39.7 & 34:54:18.6 & 15.20 & 0.61 & 45.6 & 0.25 & 3.32 \nl 
O\_1189\_216659 & NGC 653 & 01:39:31.7 & 35:23:12.3 & 14.17 & 1.42 & 83.3 & 0.36 & 2.96 \nl 
O\_1189\_199608 & 01446+3547 & 01:44:34.6 & 35:47:04.8 & 14.76 & 1.08 & 44.8 & 0.39 & 5.38 \nl 
O\_1225\_610652 & None & 01:45:28.6 & 33:35:47.0 & 16.99 & 0.52 & 48.1 & 1.81 & 2.49 \nl 
O\_1225\_389102 & UGC 1307 & 01:47:51.9 & 35:41:06.0 & 14.32 & 1.34 & 76.9 & 0.26 & 18.09 \nl 
O\_1225\_412197 & UGC 1339 & 01:49:28.8 & 35:36:36.5 & 14.39 & 1.55 & 48.8 & 0.29 & 34.59 \nl 
O\_1225\_369028 & NGC 714 & 01:50:33.0 & 35:58:31.7 & 14.06 & 1.62 & 77.5 & 0.20 & 42.61 \nl 
O\_1225\_369332 & UGC 1363 & 01:50:58.4 & 35:59:02.1 & 14.71 & 1.49 & 70.0 & 0.29 & 47.93 \nl 
O\_1225\_394977 & 01561+3549 & 01:56:09.1 & 35:49:14.2 & 14.93 & 1.48 & 79.2 & 0.12 & 9.32 \nl 
O\_1225\_484388 & UGC 1569 & 02:01:55.5 & 34:54:35.3 & 15.47 & 1.46 & 47.3 & 0.53 & 2.50 \nl 
O\_1225\_531870 & 02023+3434 & 02:02:19.7 & 34:33:02.9 & 15.01 & 1.15 & 51.8 & 0.76 & 2.67 \nl 
O\_1225\_606718 & 02087+3349 & 02:08:40.7 & 33:48:35.6 & 14.64 & 1.13 & 63.9 & 1.58 & 2.54 \nl 
\enddata
\end{deluxetable}
\clearpage

\begin{deluxetable}{llccccccr} 
\tablecaption {The Arecibo Observations \label{HI_data}}
\scriptsize
\tablenum{2}
\tablewidth{0pt} 
\tablehead{APS ID & Date\tablenotemark{a} & $V_{Sun}$ & ${\Delta}V$ & $F_{obs}$ & Spin\tablenotemark{b} & $\mu_{E-W}$ & $P_{cc}$ & Comments\\ 
 & & \multicolumn{2}{c}{(\kms)} & (Jy-\kms) &  & (\kms) & &  }
\tablenotetext{a}{Observation date is the day of the month of August 1998.  All Dates are UTC.}
\tablenotetext{b}{There are three possible values for the spin direction, ``N'' for north, ``S'' for South, and ``U'' for Undetermined. The spin direction is determined from the values of $\mu_{E-W}$ and $P_{cc}$.}
\startdata
O\_778\_873376 & 11 & 5604 & 334 &  1.03(0.08) & N & $-19.7$ & $-0.203$ &  \nl 
O\_778\_731211 & 08 & --- & --- & --- & U & --- & --- & Non-detection \nl 
O\_778\_700353 & 18 & --- & --- & --- & U & --- & --- & Non-detection \nl 
O\_778\_847676 & 16 & --- & --- & --- & U & --- & --- & Non-detection \nl 
O\_778\_849054 & 13 & --- & --- & --- & U & --- & --- & Non-detection \nl 
O\_778\_1040599 & 07 & 6379 & 328 &  1.12(0.08) & U & $+14.7$ & $-0.151$ &  \nl 
O\_778\_923367 & 14 & 6576 & 547 &  2.85(0.16) & U & $-11.4$ & $+0.389$ &  \nl 
O\_778\_755586 & 12 & 6661 & 269 &  1.23(0.10) & U & $-8.5$ & $-0.057$ &  \nl 
O\_1184\_28270 & 12+17 & 6705 & 443 &  1.30(0.10) & S & $+38.7$ & $+0.121$ &  \nl 
O\_1184\_66567 & 13 & 6791 & 267 &  0.85(0.09) & U & $-6.1$ & $+0.274$ &  \nl 
O\_1184\_128370 & 06 & 3869 & 200 &  4.55(0.06) & U & $-12.3$ & $-0.023$ &  \nl 
O\_1184\_81567 & 09 & 6588 & 283 &  1.50(0.07) & S & $+21.8$ & $+0.061$ &  \nl 
O\_1184\_196807 & 06 & 6597 & 342 &  3.10(0.09) & S & $+27.6$ & $-0.068$ &  \nl 
O\_1184\_121956 & 07 & 6436 & 358 &  9.36(0.10) & S & $+41.3$ & $-0.163$ &  \nl 
O\_1184\_275037 & 06 & 3691 & 275 &  5.25(0.09) & N & $-28.7$ & $-0.030$ &  \nl 
O\_1184\_313727 & 06 & 3674 & 224 &  7.24(0.08) & U & $+12.4$ & $-0.054$ &  \nl 
O\_1184\_347214 & 14 & 5836 & 235 &  2.34(0.07) & U & $+7.5$ & $-0.061$ &  \nl 
O\_1184\_189398 & 09 & 6838 & 406 &  1.58(0.10) & S & $+25.0$ & $+0.143$ &  \nl 
O\_843\_144830 & 15 & --- & --- & --- & U & --- & --- & Non-detection \nl 
O\_843\_65466 & 16 & 5879 & 235 &  3.24(0.09) & U & $+8.8$ & $-0.220$ &  \nl 
O\_914\_404205 & 07 & 5903 & 555 &  5.24(0.10) & N & $-74.2$ & $-0.271$ &  \nl 
O\_914\_406137 & 13+17 & 5534 & 187 &  0.28(0.09) & U & $+9.0$ & $+0.300$ &  \nl 
O\_914\_344933 & 12 & 5225 & 435 &  1.95(0.09) & S & $+29.8$ & $-0.210$ &  \nl 
O\_914\_409641 & 11 & 5466 & 213 &  4.38(0.08) & U & $+12.7$ & $+0.001$ &  \nl 
O\_914\_371708 & 16 & --- & --- & --- & U & --- & --- & Non-detection \nl 
O\_914\_413061 & 15 & --- & --- & --- & U & --- & --- & Non-detection \nl 
O\_914\_511814 & 20 & 8802 & 372 &  4.56(0.09) & S & $+24.4$ & $+0.068$ &  50 MHz bandpass\nl 
O\_914\_416774 & 14 & 5191 & 507 &  3.87(0.10) & N & $-42.9$ & $+0.079$ &  \nl 
O\_914\_286281 & 14 & 5208 & 395 &  6.34(0.09) & N & $-36.9$ & $-0.047$ &  \nl 
O\_914\_286479 & 17 & 5057 & 224 &  1.82(0.09) & U & $-11.6$ & $-0.148$ &  \nl 
O\_914\_514191 & 15 & 7134 & 261 &  3.10(0.07) & U & $-13.3$ & $+0.049$ &  \nl 
O\_914\_437214 & 15 & 6895 & 318 &  1.43(0.08) & N & $-20.7$ & $-0.162$ &  \nl 
O\_1257\_181710 & 16 & 5111 & 272 &  1.34(0.08) & U & $+2.5$ & $+0.214$ &  \nl 
O\_1257\_106449 & 07 & 4844 & 245 & 11.67(0.06) & S & $+36.4$ & $-0.014$ &  \nl 
O\_1257\_149828 & 12 & 4647 & 302 & 12.82(0.17) & N & $-44.9$ & $+0.072$ &  \nl 
O\_1257\_140283 & 13 & 4970 & 245 &  5.07(0.07) & S & $+20.6$ & $+0.029$ &  \nl 
O\_1257\_307025 & 17 & --- & --- & --- & U & --- & --- & Non-detection \nl 
O\_1257\_224112 & 18 & --- & --- & --- & U & --- & --- & Non-detection \nl 
O\_1257\_212633 & 13 & 4835 & 245 &  4.47(0.09) & N & $-24.2$ & $-0.001$ &  \nl 
O\_1244\_265500 & 15 & 6765 & 411 &  8.75(0.12) & N & $-44.2$ & $+0.051$ &  \nl 
O\_1244\_376417 & 15 & 6287 & 390 &  3.90(0.09) & U & $+14.4$ & $+0.367$ &  \nl 
O\_1244\_340721 & 11 & 6269 & 446 & 10.75(0.10) & N & $-46.2$ & $-0.234$ &  \nl 
O\_1244\_270335 & 14 & 6296 & 275 &  1.99(0.09) & N & $-24.5$ & $-0.058$ &  \nl 
O\_1244\_679996 & 06 & 4635 & 179 &  3.15(0.14) & U & $+8.9$ & $+0.093$ &  \nl 
O\_1244\_578706 & 17 & 6966 & 288 &  3.08(0.10) & U & $-4.3$ & $-0.404$ &  \nl 
O\_1244\_241809 & 12 & 4593 & 138 &  5.55(0.10) & U & $+11.4$ & $+0.164$ &  \nl 
O\_1244\_767827 & 06 & 4148 & 200 &  4.71(0.09) & S & $+18.8$ & $-0.015$ &  \nl 
O\_1244\_185805 & 18 & 6292 & 419 &  2.05(0.10) & N & $-21.7$ & $+0.069$ &  \nl 
O\_1244\_655424 & 14 & 5238 & 342 &  4.31(0.08) & N & $-51.8$ & $+0.149$ &  \nl 
O\_1244\_554203 & 16 & --- & --- & --- & U & --- & --- & Non-detection \nl 
O\_1244\_595079 & 13 & 5249 & 248 &  5.04(0.06) & N & $-30.4$ & $+0.037$ &  \nl 
\tablebreak
O\_1244\_521380 & 20 & 4891 & 283 &  1.43(0.08) & U & $-10.5$ & $+0.181$ &  \nl 
O\_1244\_442897 & 20 & --- & --- & --- & U & --- & --- & Strong RFI \nl 
O\_1244\_484644 & 18 & 4901 & 459 &  1.34(0.09) & N & $-38.6$ & $+0.220$ &  \nl 
O\_601\_2598615 & 06 & 5059 & 382 &  4.71(0.19) & N & $-52.7$ & $-0.049$ &  \nl 
O\_601\_927741 & 13 & 4575 & 323 &  3.60(0.08) & N & $-28.7$ & $-0.023$ &  \nl 
O\_601\_2448395 & 17 & 4972 & 219 &  1.69(0.08) & U & $-6.2$ & $+0.907$ &  \nl 
O\_601\_1985337 & 07 & 4892 & 227 &  2.97(0.07) & N & $-17.8$ & $-0.346$ &  \nl 
O\_601\_1986315 & 18 & --- & --- & --- & U & --- & --- & Non-detection \nl 
O\_601\_2601958 & None & --- & --- & --- & U & --- & --- &  \nl 
O\_601\_1152906 & 11 & 6267 & 446 & 10.77(0.10) & N & $-46.2$ & $-0.235$ &  \nl 
O\_601\_2454111 & None & --- & --- & --- & U & --- & --- &  \nl 
O\_601\_2363374 & 18 & --- & --- & --- & U & --- & --- & Baseline distorted \nl 
O\_601\_1044227 & 17 & 5635 & 232 &  1.13(0.09) & U & $+18.9$ & $+1.021$ &  \nl 
O\_601\_1267760 & 18 & 4623 & 318 &  2.08(0.10) & N & $-18.7$ & $-0.071$ &  \nl 
O\_601\_1050939 & 18 & 5087 & 146 &  1.42(0.07) & N & $-27.4$ & $+0.100$ &  \nl 
O\_601\_1498038 & 12 & 4758 & 523 &  9.16(0.12) & N & $-73.6$ & $-0.213$ &  \nl 
O\_601\_1063679 & 14 & 5535 & 416 &  7.15(0.10) & N & $-42.8$ & $+0.047$ &  \nl 
O\_601\_863271 & None & --- & --- & --- & U & --- & --- &  \nl 
O\_601\_1294945 & 16 & 6218 & 267 &  2.85(0.08) & N & $-15.8$ & $+0.039$ &  \nl 
O\_601\_978154 & 20 & 5823 & 347 &  1.15(0.13) & U & $+9.7$ & $+0.592$ &  \nl 
O\_601\_1197694 & 16 & 6213 & 277 &  3.72(0.08) & U & $+13.2$ & $+0.271$ &  \nl 
O\_601\_657010 & 16 & 5069 & 219 &  2.51(0.09) & S & $+18.6$ & $+0.135$ &  \nl 
O\_601\_992015 & None & --- & --- & --- & U & --- & --- &  \nl 
O\_601\_175820 & 20 & --- & --- & --- & U & --- & --- & Non-detection \nl 
O\_601\_264028 & None & --- & --- & --- & U & --- & --- &  \nl 
O\_601\_359427 & 20 & 4643 & 267 &  1.22(0.09) & U & $+3.4$ & $+0.262$ &  \nl 
O\_406\_436424 & 20 & 4708 & 221 &  1.84(0.08) & S & $+17.8$ & $+0.080$ &  \nl 
O\_406\_502375 & 16 & 4160 & 299 &  3.45(0.10) & U & $-54.1$ & $+0.958$ &  \nl 
O\_1189\_285025 & 17 & 4851 & 277 &  1.00(0.11) & U & $+0.5$ & $+0.090$ &  \nl 
O\_1189\_293769 & 06 & 4839 & 534 &  5.36(0.10) & N & $-70.2$ & $-0.008$ &  \nl 
O\_1189\_251296 & 12 & 4127 & 213 &  3.28(0.07) & U & $+8.9$ & $+0.104$ &  \nl 
O\_1189\_224928 & 13 & 4884 & 497 &  6.83(0.10) & S & $+74.1$ & $-0.032$ &  \nl 
O\_1189\_244311 & 14 & --- & --- & --- & U & --- & --- & Strong RFI \nl 
O\_1189\_234487 & 15 & 5126 & 176 &  1.71(0.08) & U & $-6.4$ & $+0.164$ &  \nl 
O\_1189\_216659 & 16 & --- & --- & --- & U & --- & --- & Strong RFI \nl 
O\_1189\_199608 & 17 & 4786 & 272 &  1.11(0.08) & N & $-19.7$ & $+0.376$ &  \nl 
O\_1225\_610652 & 09 & 5728 & 221 &  1.11(0.07) & U & $-12.0$ & $-0.104$ &  \nl 
O\_1225\_389102 & 18 & --- & --- & --- & U & --- & --- & Non-detection \nl 
O\_1225\_412197 & 18 & --- & --- & --- & U & --- & --- & Non-detection \nl 
O\_1225\_369028 & None & --- & --- & --- & U & --- & --- &  \nl 
O\_1225\_369332 & 20 & --- & --- & --- & U & --- & --- & Non-detection \nl 
O\_1225\_394977 & 12 & 5406 & 366 &  1.75(0.08) & S & $+34.6$ & $+0.099$ &  \nl 
O\_1225\_484388 & 13 & --- & --- & --- & U & --- & --- & Non-detection \nl 
O\_1225\_531870 & 10 & 4334 & 288 &  1.14(0.09) & U & $+13.2$ & $-0.211$ &  \nl 
O\_1225\_606718 & 08 & 6128 & 435 &  3.33(0.10) & N & $-39.1$ & $+0.326$ &  \nl 
\enddata
\end{deluxetable}
\clearpage

\begin{deluxetable}{lrrrrrr} 
\tablecaption{Arecibo $\vec{L}$ Anisotropy Probabilities, $P(V)$ \label{LvecTests}}
\scriptsize
\tablenum{3}
\tablewidth{0pt} 
\tablehead{  & \multicolumn{3}{c}{Arecibo Sample} & \multicolumn{3}{c}{MAPS-PP Catalog}\\ 
Distribution & all & hi dens & lo dens & all & hi dens & lo dens}
\startdata
$N_{gal}$ & 54 & 23 & 30 & 1230 & 616 & 615 \nl
$\theta_{\vec{L}}$ & 0.220 & 0.061 & 0.064 & --- & --- & --- \nl
$\theta$ & 0.449 & 0.448 & 0.807 & 0.388 & 0.270 & 0.341 \nl
$\Delta\theta_{\vec{L}}(1)$ &0.096 & 0.462 & 0.120 & --- & --- & --- \nl
$\Delta\theta(1)$ & 0.121 & 0.119 & 0.361 & 0.667 & 0.197 & 0.352 \nl
$\Delta\theta(Geo)$ & --- & --- & --- & 0.606 & 0.730 & 0.680 \nl
$\Delta\theta_{\vec{L}}(Geo)$ & 0.384 & 0.894 & 0.374 & --- & --- & --- \nl
$\Delta\theta(Ridge)$ & --- & --- & --- & 0.799 & 0.126 & 0.900 \nl
$\Delta\theta_{\vec{L}}(Ridge)$ & 0.257 & 0.142 & 0.113 & --- & --- & --- \\[1.5ex]
\tableline\tablevspace{1.5ex}
\multicolumn{7}{c}{``$\Delta\theta_{\vec{L}}$/$\Delta\theta$'' represents $\Delta\theta_{\vec{L}}$ for Arecibo Sample and $\Delta\theta$ for MAPS-PP Catalog,} \\
\multicolumn{7}{c}{``(N)'' below indicates the number of galaxies in this distribution.} \\[1.5ex]
\tableline\tablevspace{1.5ex}
$\Delta\theta_{\vec{L}}$/$\Delta\theta(GC-10)$ & --- (000) & --- (000) & --- (000) & 0.651 (013) & 0.651 (013) & --- (000) \nl
$\Delta\theta_{\vec{L}}$/$\Delta\theta(GC-15)$ & 0.279 (001) & 0.279 (001) & --- (000) & 0.700 (093) & 0.700 (093) & --- (000) \nl
$\Delta\theta_{\vec{L}}$/$\Delta\theta(GC-20)$ & 0.699 (009) & 0.884 (008) & 0.279 (001) & 0.823 (224) & 0.846 (220) & 0.731 (004) \nl
$\Delta\theta_{\vec{L}}$/$\Delta\theta(GC-25)$ & 0.489 (018) & 0.954 (014) & 0.108 (003) & 0.113 (431) & 0.062 (389) & 0.797 (043) \nl
$\Delta\theta_{\vec{L}}$/$\Delta\theta(GC-30)$ & 0.086 (030) & 0.410 (021) & 0.252 (008) & 0.774 (581) & 0.727 (455) & 0.596 (127) \nl
$\Delta\theta_{\vec{L}}$/$\Delta\theta(GC-35)$ & 0.046 (042) & 0.065 (023) & 0.160 (018) & 0.590 (722) & 0.881 (497) & 0.639 (226) \nl
$\Delta\theta_{\vec{L}}$/$\Delta\theta(GC-40)$ & 0.124 (048) & 0.244 (023) & 0.624 (024) & 0.095 (817) & 0.113 (508) & 0.946 (310) \nl
$\Delta\theta_{\vec{L}}$/$\Delta\theta(GC-45)$ & 0.214 (052) & 0.016 (023) & 0.815 (028) & 0.307 (902) & 0.380 (509) & 0.692 (394) \nl
$\Delta\theta_{\vec{L}}$/$\Delta\theta(GC-50)$ & 0.494 (052) & 0.106 (023) & 0.917 (028) & 0.616 (948) & 0.742 (509) & 0.905 (440) \nl
$\Delta\theta_{\vec{L}}$/$\Delta\theta(GCR-10)$ & --- (000) & --- (000) & --- (000) & 0.858 (013) & 0.858 (013) & --- (000) \nl
$\Delta\theta_{\vec{L}}$/$\Delta\theta(GCR-15)$ & 0.279 (001) & 0.279 (001) & --- (000) & 0.839 (093) & 0.839 (093) & --- (000) \nl
$\Delta\theta_{\vec{L}}$/$\Delta\theta(GCR-20)$ & 0.931 (009) & 0.683 (008) & 0.279 (001) & 0.741 (224) & 0.720 (220) & 0.232 (004) \nl
$\Delta\theta_{\vec{L}}$/$\Delta\theta(GCR-25)$ & 0.764 (018) & 0.365 (014) & 0.850 (003) & 0.779 (431) & 0.665 (389) & 0.862 (043) \nl
$\Delta\theta_{\vec{L}}$/$\Delta\theta(GCR-30)$ & 0.767 (030) & 0.523 (021) & 0.871 (008) & 0.346 (581) & 0.605 (455) & 0.765 (127) \nl
$\Delta\theta_{\vec{L}}$/$\Delta\theta(GCR-35)$ & 0.359 (042) & 0.473 (023) & 0.707 (018) & 0.040 (722) & 0.247 (497) & 0.546 (226) \nl
$\Delta\theta_{\vec{L}}$/$\Delta\theta(GCR-40)$ & 0.917 (048) & 0.711 (023) & 0.475 (024) & 0.670 (817) & 0.264 (508) & 0.271 (310) \nl
$\Delta\theta_{\vec{L}}$/$\Delta\theta(GCR-45)$ & 0.281 (052) & 0.253 (023) & 0.428 (028) & 0.457 (902) & 0.849 (509) & 0.127 (394) \nl
$\Delta\theta_{\vec{L}}$/$\Delta\theta(GCR-50)$ & 0.148 (052) & 0.568 (023) & 0.039 (028) & 0.274 (948) & 0.196 (509) & 0.973 (440) \nl 
\enddata
\end{deluxetable}
\clearpage

%
%

%
%

\clearpage
\figcaption[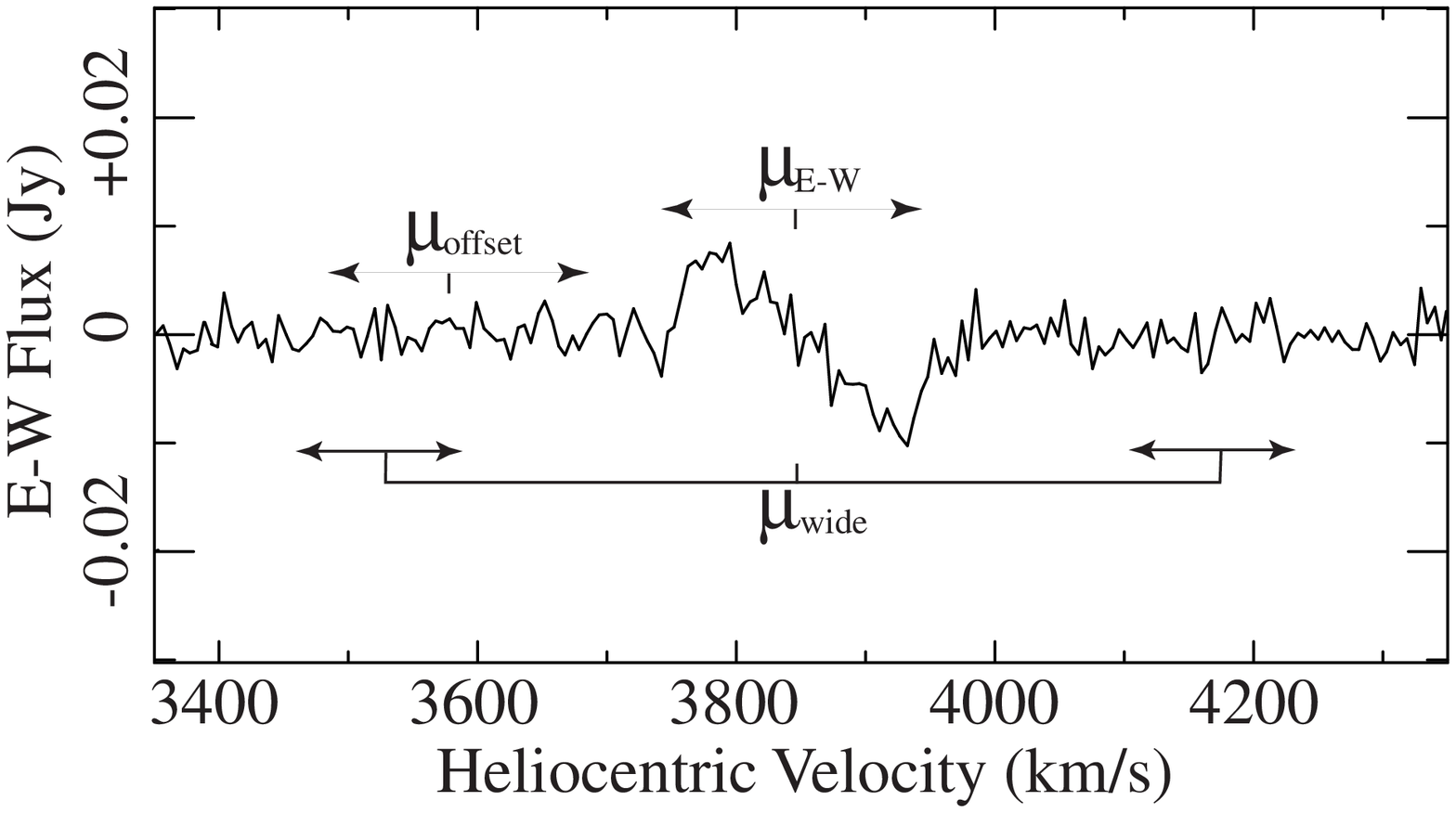]{This plot shows the ranges of velocity over which
 $\mu_{E-W}$, $\mu_{offset}$, and $\mu_{wide}$ are determined, in this case for
the spectrum of UGC 12231.  \label{mudef}}

\figcaption[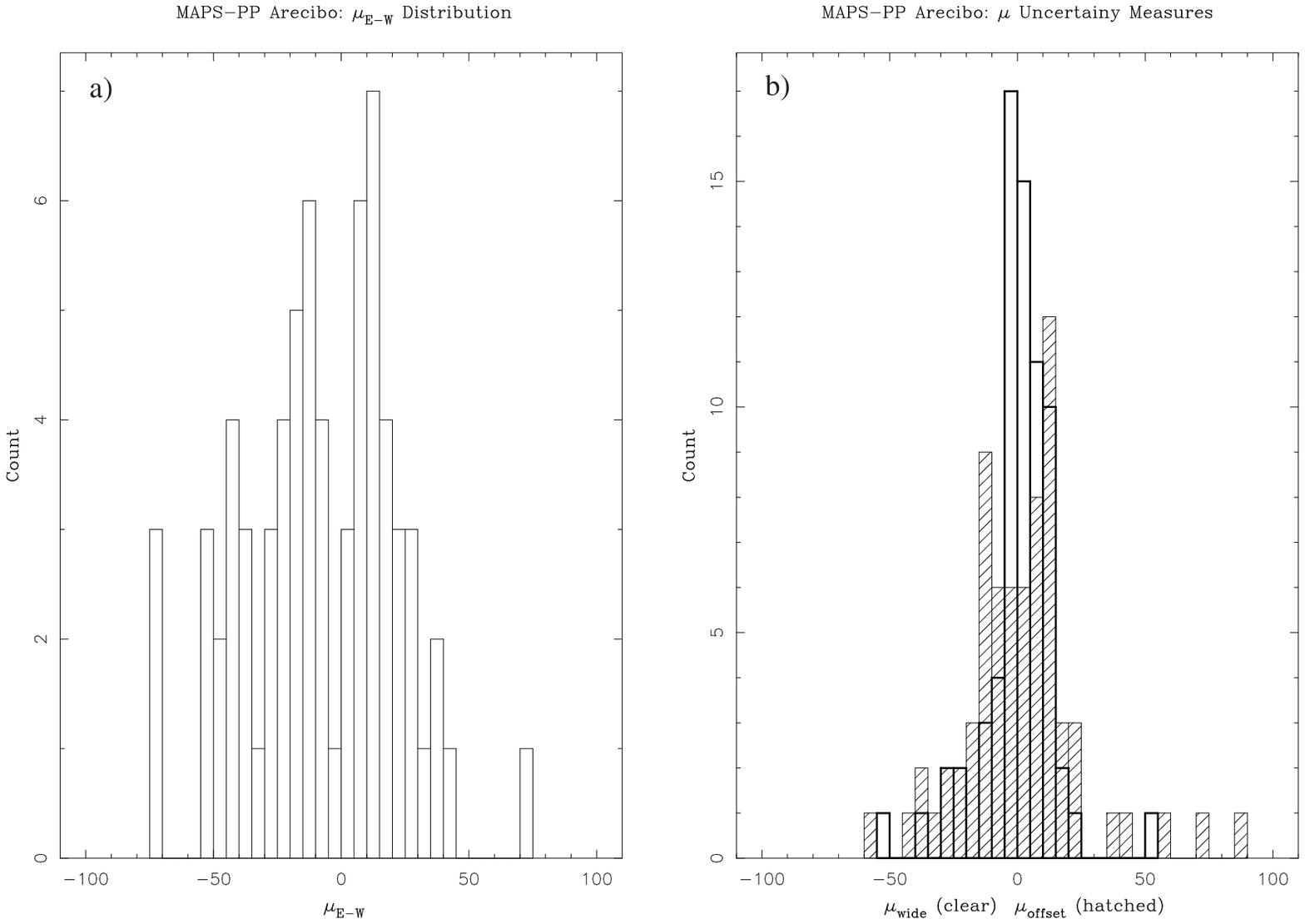]{The plot on the left shows
the distribution of $\mu_{E-W}$ values for the 70 Arecibo sample galaxies
detected in {\HI}.  Negative $\mu_{E-W}$ means the galaxy's $\vec{L}$ points
northward.  The right plot shows the distribution of $\mu_{offset}$ (hatched)
and $\mu_{wide}$ (clear), both of which are have a FWHM of roughly 15 {\kms}.
\label{fig3-1}}

\figcaption[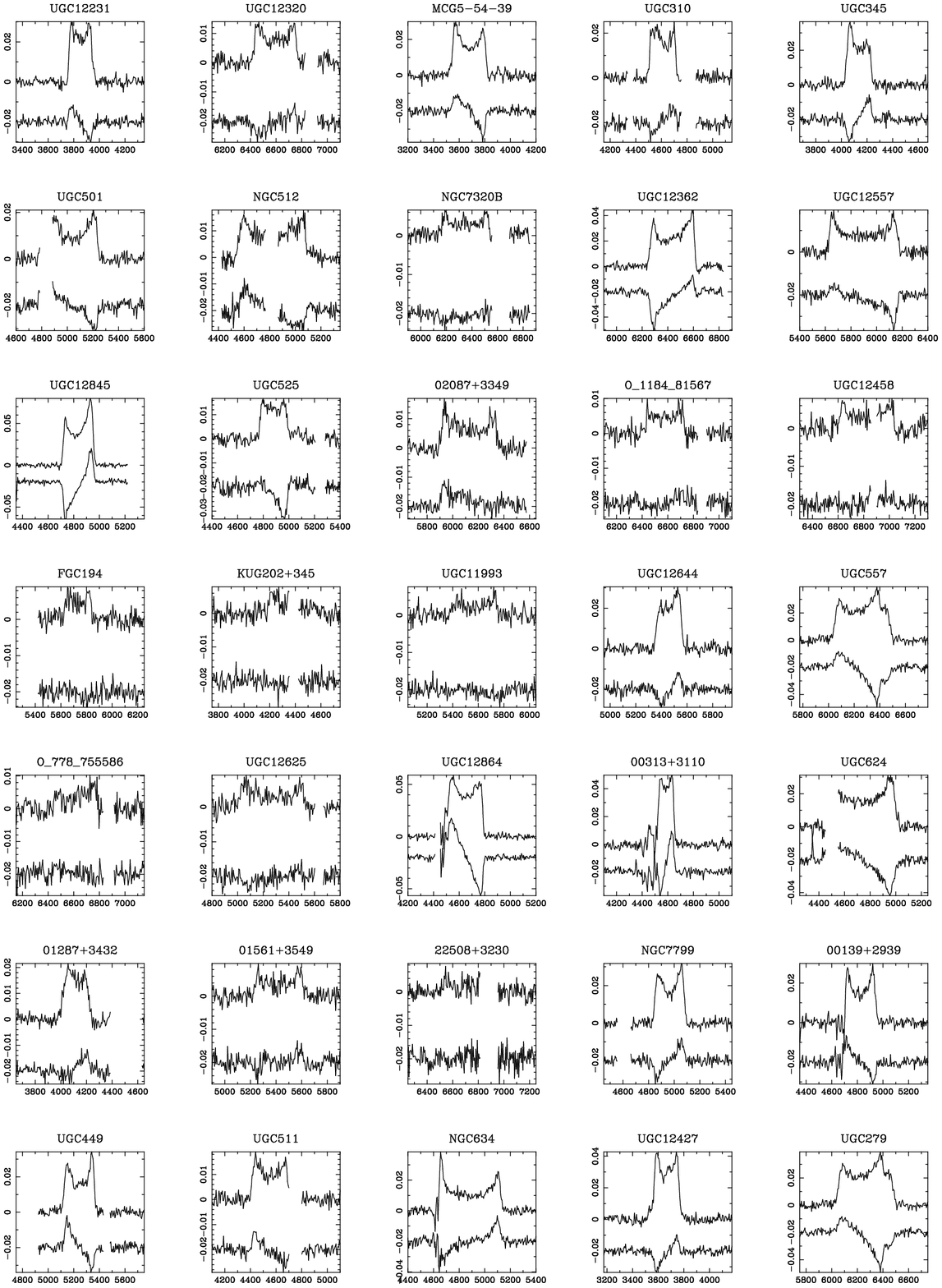]{The reduced spectra for
the Arecibo sample in units of Jy versus {\kms}. The top spectrum in each pair
is $f_{E+W}(v)$, the bottom spectrum is $f_{E-W}(v)$ shifted by -0.02 Jy.  Each
spectrum is 1500 {\kms} wide and centered on the systemic heliocentric velocity
of the galaxy.  Spectra are baseline corrected.  Gaps indicate where RFI was
identified.  \label{spectra}}

\figcaption[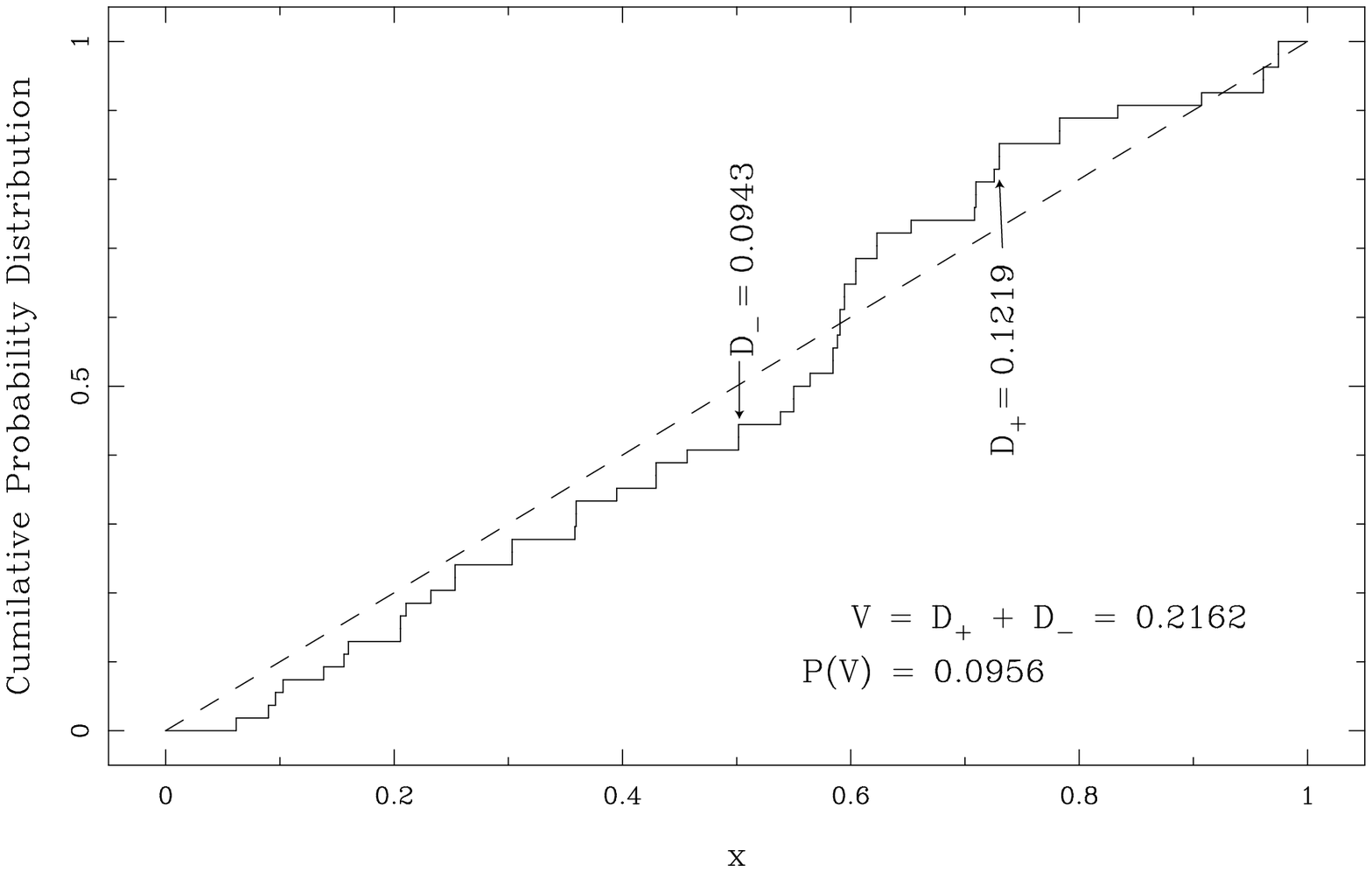]{This diagram illustrates how the Kuiper V
statistic is determined for a given cumulative distribution.  The dashed line
is the cumulative distribution, $S(x)$, of $\Delta\theta_{\vec{L}}(1)$ for the
Arecibo sample.The solid line is the modeled, isotropic distribution, $S_m(x)$.
 The maximum positive ($D_{+}$) and negative ($D_{-}$) differences between
$S(x)$ and $S_m(x)$ are shown for this sample.  The Kuiper $V$ statistic is the
sum of $D_{+}$ and $D_{-}$ whereas the K--S $D$ statistic is $D_{+}$ (because
$|D_{+}|>|D_{+}|$ in this case).  It has been previously established that $V$
is a more robust measure than $D$ of the difference between two distributions
and thus we choose to use $V$ to measure the anisotropy of our data (see Press
{\etal} 1992, for example). \label{VDef}}

\figcaption[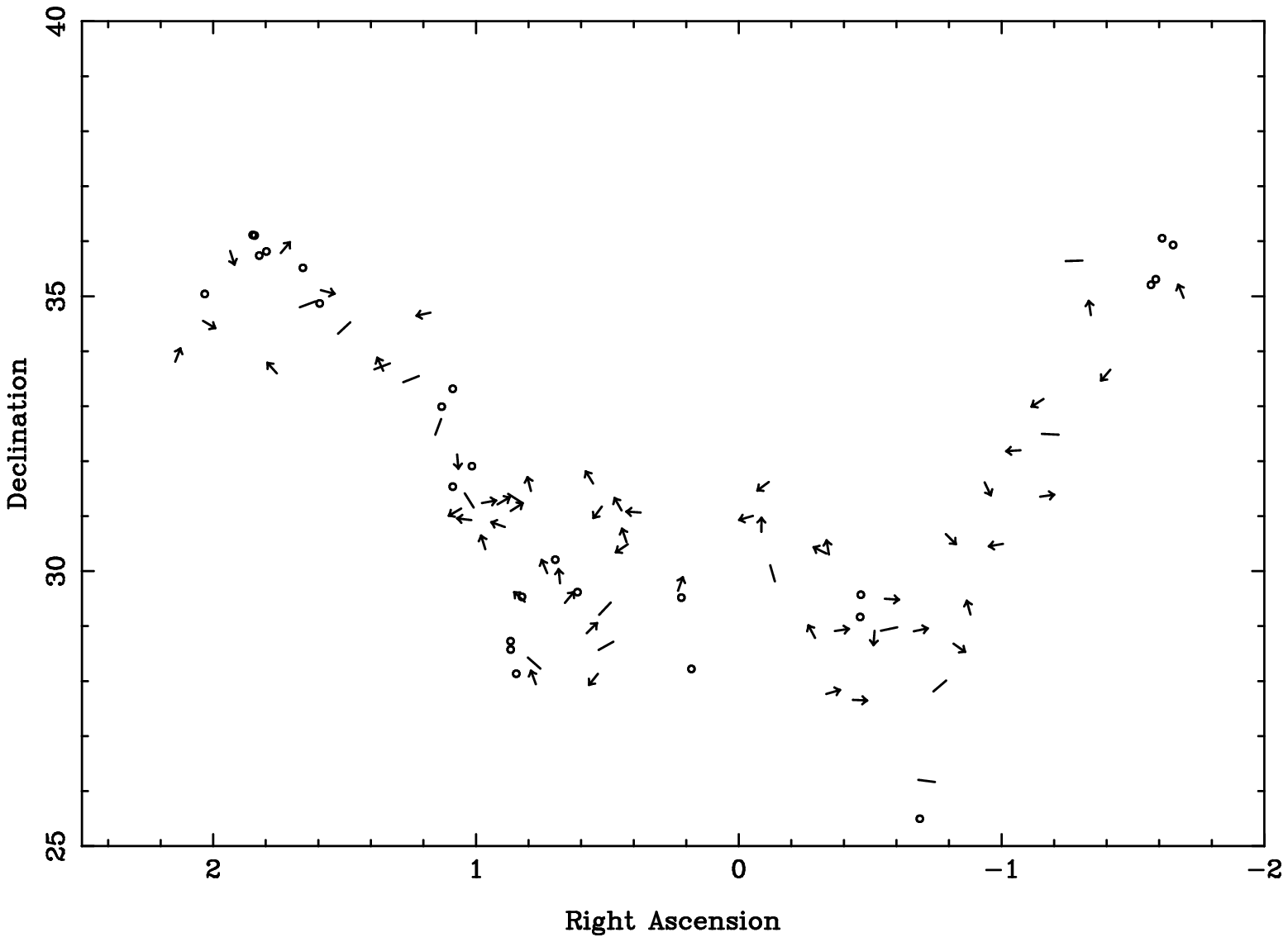]{This map shows the distribution of the entire
Arecibo subsample on the sky, with varying symbols depending on the value of
$\theta_{\vec{L}}$.  If $\theta_{\vec{L}}$ is well-determined, an arrow shows
its direction, if $\theta_{\vec{L}}$ is not-well determined, but the galaxy was
detected in {\HI}, a line shows the direction of $\theta_{\vec{L}}$, but not
the sign.  A circle marks those galaxies that were undetected in {\HI}.
\label{lvecmap}}

\figcaption[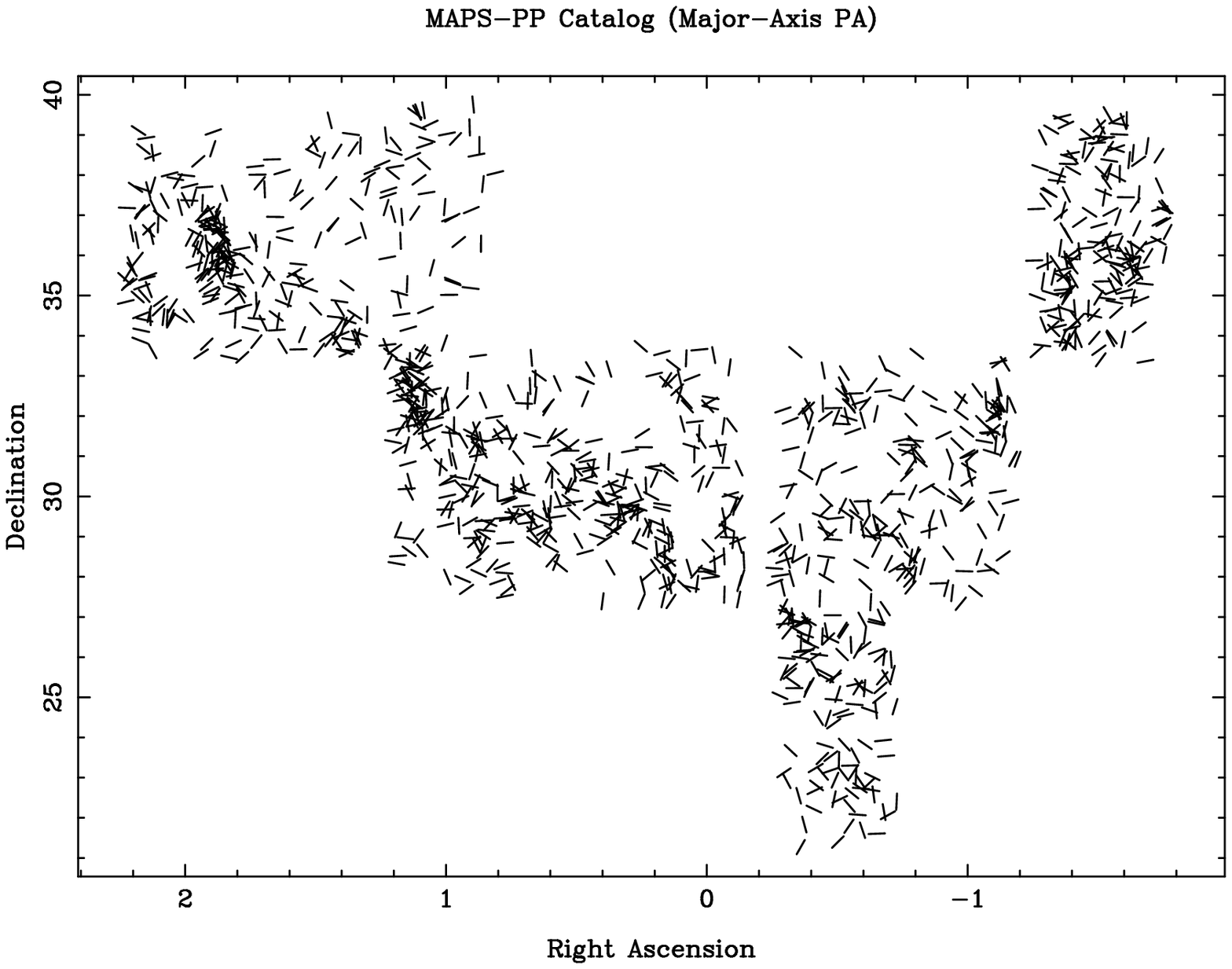]{This map shows the major-axis position angle
distribution of MAPS-PP O sample of 1230 galaxies on the sky.  Note that
apparent alignments are visible to the eye. \label{mapspppa}}

\figcaption[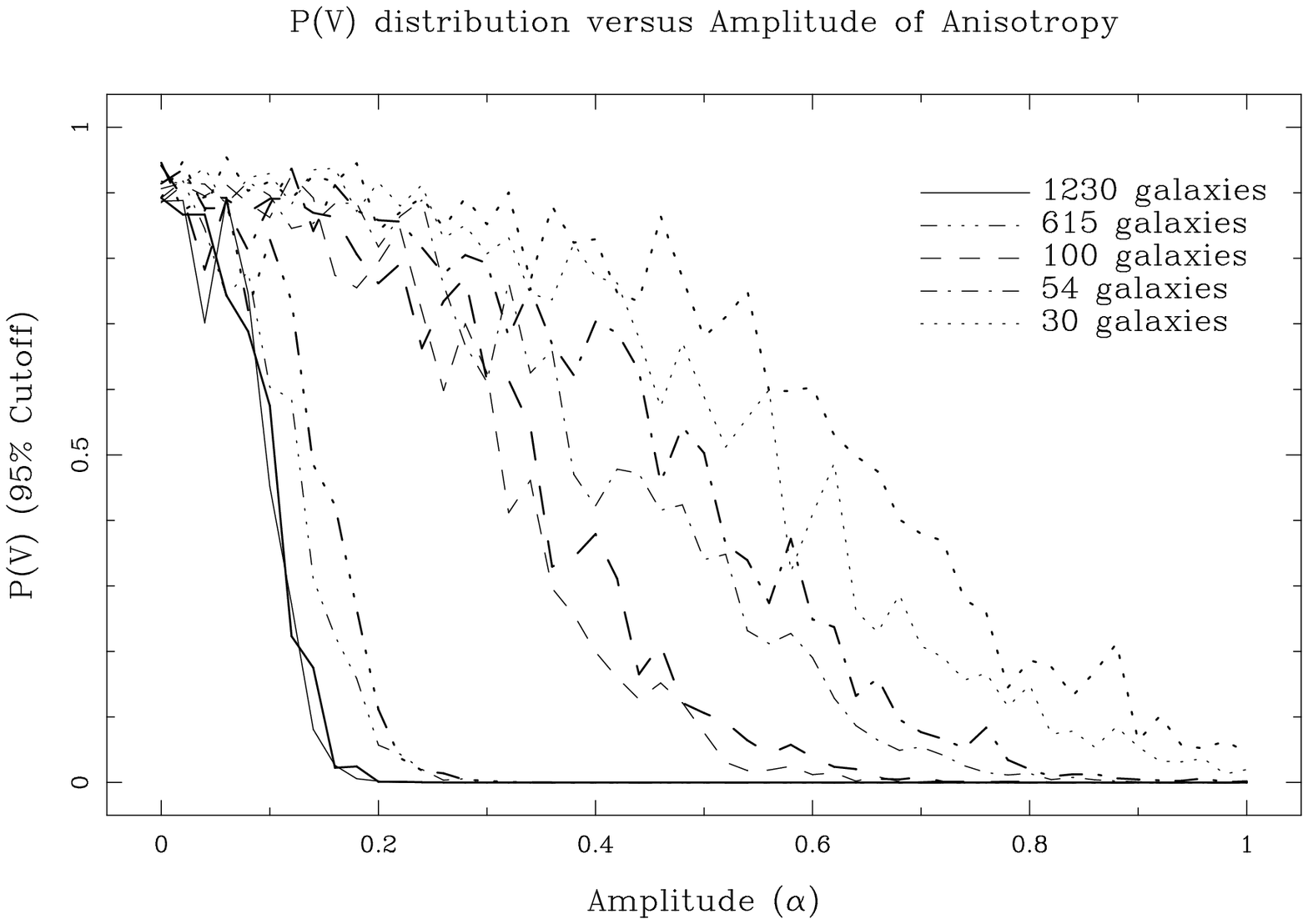]{This plot shows the 95\% cutoff value of
$P(V)$, which is the value of $P(V)$ which 95\% of all distributions lie below.
 Samples of 30, 54, 100, 615, and 1230 galaxies were generated via equation
\ref{cos_prob} (thin lines) and \ref{cos_prob_2} (thick lines).   The value of
$\alpha$ for which these lines drop below a value of 0.05 is referred to as
$\alpha_{95}$.  $\alpha_{95}$ represents the smallest amplitude of a sinusoidal
anisotropy for which 95\% of the distributions would be detected via the Kuiper
Test.  Notice the similar results for distributions generated by both equation
\ref{cos_prob} and equation \ref{cos_prob_2}. \label{ampVprobs}}

\end{document}